\documentclass[a4paper]{jpconf}
\usepackage{graphicx}
\usepackage{slashed}
\usepackage{amsmath}
\usepackage{amsfonts}
\usepackage{amssymb}
\usepackage{wrapfig}
\usepackage{hyperref}
\usepackage{mathrsfs}
\usepackage{color}
\usepackage{subfig}

\begin{document}

\begin{flushright}
WITS-MITP-012 
\end{flushright}

\title{Single Top and Higgs Production in  $e^- p$ collisions }

\author{Mukesh Kumar}

\address{National Institute for Theoretical Physics, School of Physics,
                School of Physics and Mandelstam Institute for Theoretical Physics, University of the Witwatersrand, Johannesburg, Wits 2050, South Africa.}

\ead{mukesh.kumar@cern.ch}

\begin{abstract}
In this proceedings some studies on the prospects of single top production at 
the Large Hadron Electron Collider (LHeC) and double Higgs production at the
Future Circular Hadron Electron Collider (FCC-he) shall be presented.
In particular, we investigated the $tbW$ couplings via single top quark
production with the introduction of possible anomalous Lorentz structures,
and measured the sensitivity of the Higgs self coupling ($\lambda$) through 
double Higgs production. The studies are performed with 60 GeV electrons 
colliding with 7 (50) TeV protons for the LHeC (FCC-he).
 
For the single top studies a parton level study has been performed, and we find the
sensitivity of the anomalous coupling at a 95\% C.L, considering 10-1\% 
systematic errors. The double Higgs production has been studied with speculated
detector parameters and the sensitivity of $\lambda$ estimated via the cross
section study around the Standard Model Higgs self coupling strength
($\lambda_{SM}$) considering 5\% systematic error in signal and backgrounds.
Effects of non-standard CP-even and CP-odd couplings for $hhh$, $hWW$ and
$hhWW$ vertices have been studied and constrained at 95\% C.L.
\end{abstract}

\section{Introduction}
\label{intro}
In the Standard Model (SM) of particle physics the top quark, the heaviest
among all matter particles, and the Higgs boson, a particle responsible for
giving masses to all matter particles and gauge bosons, play a crucial role
in the model. And hence it has been, and still is, a challenge for colliders 
to study different characteristics of these two particles. These characteristics 
include their charge, mass, interactions and coupling strengths with other 
particles etc.

As such we shall briefly review some of the important properties of the top
quark and the Higgs boson from theoretical calculations and results from present
and past colliders. We will then present some predictions for a future $e^- p$ 
collider known as the Large Hadron Electron Collider (LHeC) and the 
Future Circular Hadron Electron Collider (FCC-he) at a center of mass energy 
$\sqrt{s} \approx 1.3$ TeV and $\sqrt{s} \approx 3.5$ TeV, respectively.

The structure of this proceeding shall then be: section \ref{top} is devoted to 
the top quark studies, sections \ref{Higgslhec} and \ref{Higgsfcc} are devoted 
for the Higgs boson studies in $e^-p$ collisions. We conclude our
inferences based on our studies in section \ref{conc}.

\section{The top quark at the LHeC}
\label{top}
In this section we briefly review the physics potential of the proposed
LHeC by estimating the accuracy of anomalous
$Wtb$ couplings in the single anti-top quark production through $e^- p$
collisions~\cite{Dutta:2013mva}. Within the SM the $Wtb$ vertex is purely
left-handed. However, the most general lowest dimension CP conserving (in
effect of which the couplings are real) Lagrangian for this vertex is given by
\begin{equation}
{\cal L}_{Wtb}
=
\frac{g}{\sqrt 2} \left[
                    W_\mu \bar t \gamma^\mu \left( V_{tb} f_1^L P_L + f_1^R P_R \right)b
                    - \frac{1}{2 m_W} W_{\mu\nu} \bar t \sigma^{\mu\nu}\left(f_2^L P_L + f_2^R P_R \right)b
                  \right]
+ {\rm h.c}.
\end{equation}
Here $f_1^L = 1 + \Delta f_1^L$, $W_{\mu\nu} = \partial_\mu W_\nu - \partial_\nu W_\mu$,
$P_{L,R} = \frac{1}{2}\left(1 \mp \gamma_5\right)$ are left- and right-handed projection
operators, $\sigma^{\mu\nu} = {\rm i} \left(\gamma^\mu \gamma^\nu - \gamma^\nu \gamma^\mu \right)/2$
and $g=e/{\sin \theta_W}$. In the SM $\left| V_{tb}\right| f_1^L \approx 1$. So, along
with $\Delta f_1^L$ all other couplings $f_2^L, f_1^R, f_2^R$ vanish at tree-level, but
are non-zero at higher orders. The constraints on these couplings are the following:

\begin{itemize}
\item Assuming only one anomalous coupling to be non-zero at a time:
$-0.13 \leq\vert V_{tb}\vert f_1^L \leq 0.03$, $-0.0007 \leq f_1^R \leq 0.0025$,
$-0.0015 \leq f_2^L \leq 0.0004$, $-0.15 \leq f_2^R \leq 0.57$ from $B$ decays;

\item Single top production at D\O\, assuming $\vert V_{tb}\vert f_1^L =1$:
$\vert f_1^R \vert \leq 0.548$, $\vert f_2^L \vert \leq 0.224$,
$\vert f_2^R \vert \leq 0.347$;

\item Associated $tW$ production at LHC through $\gamma p$ collision:
$\vert f_1^R \vert \leq 0.55$, $\vert f_2^L \vert \leq 0.22$, $\vert f_2^R \vert \leq 0.35$;

\item ATLAS: asymmetries associated through angular distribution
Re$\left(f_1^R \right) \in$ [-0.44,0.48], Re$\left(f_2^L \right) \in$
[-0.24,0.21], Re$\left(f_2^R \right) \in$ [-0.49,0.15].
\end{itemize}
Loop Corrections:
\begin{itemize}
 \item QCD: $f_2^R=-6.61 \times 10^{-3}$,  $f_2^L=-1.118 \times 10^{-3}$ ($m_t=171$ GeV);
 \item EW: $f_2^R=-(1.24\pm1.23{\rm i}) \times 10^{-3}$, $f_2^L=-(0.102\pm0.014{\rm i}) \times 10^{-3}$ ($m_h=126$ GeV);
 \item SM: $f_2^R=-(7.85\pm1.23{\rm i}) \times 10^{-3}$, $f_2^L=-(1.220\pm0.014{\rm i}) \times 10^{-3}$.
\end{itemize}

\subsection{Single anti-top production}
\label{stopprod}
We analyze the anti-top production through the hadronic and leptonic decay modes
of $W$'s as (a) $e^- p \to \bar t \nu_e, \left( \bar t \to W^- \bar b, W^- \to jj\right)$
and (b) $e^- p \to \bar t \nu_e, \left( \bar t \to W^- \bar b, W^- \to l \nu_l \right)$,
respectively. We impose the following selection cuts on events:
\begin{itemize}
\item Minimum transverse momenta: $p_{T_{b,j}} \ge 20$ GeV, $p_{T_{b,\bar l}} \ge 25$ GeV
and $\slashed E_T \ge 25$ GeV;

\item Pseudorapidities: $\left|\eta_{\bar b,l}\right|\le 2.5$ and $\left|\eta_{j}\right|\le 2.5$;

\item Isolation cuts: $\Delta R_{ij} \ge 0.4$ where $i,j$ are leptons, light-jets and $b$-jets;

\item $\Delta \phi_{\slashed E_T,j} >$ 0.4, $\Delta \phi_{\slashed E_T,l} >$ 0.4,
$\Delta \phi_{\slashed E_T,b} >$ 0.4 and

\item $\left|m_{j_1 j_2}-m_W\right|\le 22$ GeV for the hadronic channel.
\end{itemize}
After estimation of all possible backgrounds in both channels, imposing the above
selection cuts, we observed high yields of single anti-top quark production with
fiducial efficiency of $\sim 70$ \% and $\sim 90$ \% in the hadronic and leptonic
decay modes of $W^-$ respectively.

\subsection{Estimators and $\chi^2$ analysis}
To find the sensitivity of all non-standard couplings, we follow three different
approaches based on one dimensional histograms. In the histograms we compare the
SM distributions, including all backgrounds, with all non-standard couplings
with representative values $f_1^R = + 0.5$, $f_2^L = - 0.5$, $f_2^L = + 0.5$ and
$f_2^R = + 0.5$. For hadronic modes, we consider six different distributions, namely,
$\Delta \phi _{E_T^{miss}j_1}$, $\Delta \phi _{E_T^{miss}b}$, $\Delta \phi _{E_T^{miss}W}$,
$\Delta \phi _{bW}$, $\cos \theta_{bj_1}$ and $\Delta \eta_{bj_1}$, while for
leptonic modes, there are four different one dimensional histograms used for the
analysis $\Delta \phi _{E_T^{miss}l}$, $\Delta \phi _{E_T^{miss}b}$, $\cos \theta_{bl}$
and $\Delta \eta_{bl}$.  

\subsubsection{Angular Asymmetries from Histograms:}
As a preliminary study in order to get a feel for different chiral and momentum dependencies
of couplings, the following asymmetries are defined:
\begin{eqnarray}
&A_{\theta_{ij}}&=\,\frac{ N_+^A(\cos\theta_{ij} > 0) - N_-^A(\cos\theta_{ij} < 0) }{ N_+^A(\cos\theta_{ij} > 0) + N_-^A(\cos\theta_{ij} < 0) }, \label{atheta} \\
&A_{\Delta \eta_{ij}}&=\,\frac{ N_+^A(\Delta \eta_{ij} > 0) - N_-^A(\Delta \eta_{ij} < 0) }{ N_+^A(\Delta \eta_{ij} > 0) + N_-^A(\Delta \eta_{ij} < 0) }, \label{ady} \\
&A_{\Delta\Phi_{ij}} &=\, \frac{N_+^A \left( \Delta\phi_{ij}>\frac{\pi}{2}\right)-N_-^A\left( \Delta\phi_{ij}<\frac{\pi}{2}\right)}{N_+^A\left( \Delta\phi_{ij}>\frac{\pi}{2}\right)+N_-^A\left( \Delta\phi_{ij}<\frac{\pi}{2}\right)}, \label{aphi}
\label{adphi}
\end{eqnarray}
with $0\leq\Delta\phi_{ij}\leq \pi$. The asymmetry $A_{\alpha}$ and its
statistical error for $N_{+}^A$ and $N_{-}^A$ events, where $N=\left(N_{+}^A
+N_{-}^A\right)= L \cdot\sigma$, are calculated using the following definition
based on binomial distributions:
\begin{eqnarray}
A_{\alpha} &=& a \pm \sigma_a, \qquad {\rm where}\\
a &=& \frac{N_{+}^A - N_{-}^A}{N_{+}^A+ N_{-}^A} \quad {\rm and}
\quad \sigma_a = \sqrt{\frac{1-a^2}{L\cdot\sigma}}\,;
\quad \left(\alpha=\cos\theta_{ij},\Delta \eta_{ij},\Delta \Phi_{ij} \right). \label{aerr}
\end{eqnarray}
Here $\sigma \equiv\sigma (e^-p\to \bar t \nu, \, \bar t \to W^- \bar b) \times
BR( W^- \to jj/\, l^-\bar\nu)\times \epsilon_b $ is the total cross section in
the respective channels after imposing selection cuts and $\epsilon_b=0.6$ is
the $b/\bar b$ tagging efficiency.

\begin{table}[tb]\footnotesize
\centering
\begin{tabular*}{\textwidth}{c@{\extracolsep{\fill}} cccccc}\hline\hline
& $A_{\Delta \Phi_{\slashed E_T j_1}}$ & $A_{\Delta \Phi_{\slashed E_T \bar b}}$ & $A_{\Delta \Phi_{\slashed E_T W^-}}$ & $A_{\Delta \Phi_{W^-\bar b}}$ & $A_{\theta_{\bar bj_1}}$ & $A_{\Delta \eta_{\bar b j_1}}$  \\ \hline 
SM+$\sum_i {\rm Bkg}_i$                & .532 $\pm$ .003 & .282 $\pm$ .005 & .503 $\pm$ .004 & .799 $\pm$ .003 &  .023  $\pm$ .001   & -.712 $\pm$ .003\\  
$f_{1}^{R} = +.5$ & .327 $\pm$ .004 & .231 $\pm$ .004 & .564 $\pm$ .004 & .778 $\pm$ .003 &  .0005 $\pm$ .004   & -.806 $\pm$ .003 \\  
$f_{2}^{L} = -.5$ & .528 $\pm$ .004 & .082 $\pm$ .004 & .716 $\pm$ .003 & .748 $\pm$ .003 & -.196  $\pm$ .004   & -.868 $\pm$ .002 \\  
$f_{2}^{L} = +.5$ & .390 $\pm$ .005 & .269 $\pm$ .004 & .585 $\pm$ .004 & .683 $\pm$ .004 &  .106  $\pm$ .005   & -.795 $\pm$ .003\\ 
$f_{2}^{R} = +.5$ & .330 $\pm$ .004 & .363 $\pm$ .004 & .566 $\pm$ .003 & .656 $\pm$ .003 & -.197  $\pm$ .004   & -.823 $\pm$ .002\\ \hline \hline
\end{tabular*}
\caption{\small {Asymmetries and errors associated with the kinematic
distributions in hadronic histograms at an integrated  luminosity $L$ = 100 fb$^{-1}$.
}}
\label{asym_had}
\end{table}

\begin{table}[tb]\footnotesize
\centering
\begin{tabular*}{\textwidth}{c@{\extracolsep{\fill}} ccccc}\hline\hline\hline
& $A_{\Delta \Phi_{\slashed E_T l_1}}$ & $A_{\Delta \Phi_{\slashed E_T \bar b}}$ & $A_{\theta_{\bar bl_1}}$ & $A_{\Delta \eta_{\bar b  l_1}}$   \\ \hline 
SM + $\sum_i {\rm Bkg}_i$               & .384 $\pm$ .004  & .710  $\pm$ .003 & .551 $\pm$ .006 & -.765 $\pm$ .007 \\
$f_{1}^{R} = +.5$ & .484 $\pm$ .004  & .702  $\pm$ .003 & .332 $\pm$ .006 & -.821 $\pm$ .003 \\
$f_{2}^{L} = -.5$ & .526 $\pm$ .004  & .620  $\pm$ .003 & .410 $\pm$ .006 & -.831 $\pm$ .002 \\
$f_{2}^{L} = +.5$ & .353 $\pm$ .005  & .812  $\pm$ .003 & .392 $\pm$ .007 & -.850 $\pm$ .003 \\
$f_{2}^{R} = +.5$ & .424 $\pm$ .004  & .684  $\pm$ .003 & .507 $\pm$ .005 & -.809 $\pm$ .003 
\\ \hline \hline
\end{tabular*}
\caption{\small {Asymmetries and errors associated with the kinematic
distributions in leptonic histograms at an integrated luminosity $L$ = 100 fb$^{-1}$.
}}
\label{asym_lep}
\end{table}
Tables \ref{asym_had} and \ref{asym_lep} show how asymmetries are affected due
to anomalous couplings of the order $10^{-1}$. The asymmetry suggests that the
distribution of the cosine of the angle between the tagged $\bar b$ quark and the
highest $p_T$ jet $j_1$ in the hadronic mode is the most sensitive observable.

\subsubsection{Exclusion contour from bin analysis:}
To make the analysis more effective we perform the $\chi^2$ analysis defined as:
\begin{eqnarray}
 \chi^2\left(f_i,f_j\right) 
 &=&\sum_{k=1}^N \,
  \left(\, \frac{{\cal N}^{\rm exp}_k - {\cal N}^{\rm
   th}_k\left(f_i,f_j\right) }{\delta{\cal N}^{\rm exp}_k} \,
  \right)^2, \label{chidef} 
\end{eqnarray}
where ${\cal N}_k^{\rm th}\left(f_i,f_j\right)$ and ${\cal N}_k^{\rm exp}$ are
the total number of events predicted by the theory involving $f_i,\, f_j$ and
measured in the experiment for the $k^{\rm th}$ bin. $\delta {\cal N}_k^{\rm exp}$
is the combined statistical and systematic error $\delta_{\rm sys}$ in measuring
the events for the $k^{\rm th}$ bin. If all the coefficient $f_i$'s are small,
then the experimental result in the $k^{\rm th}$ bin should be approximated by
the SM and background prediction as
\begin{eqnarray}
 {\cal N}^{\rm exp}_k \approx 
  {\cal N}^{\rm SM}_k+\sum_i {\cal N}^{\rm Bkg_i}_k= {\cal N}^{\rm SM+\sum_i Bkg_i}_k.
  \label{expsm}
\end{eqnarray}
The error $\delta {\cal N}_k^{\rm SM}$ can be defined as
\begin{eqnarray}
\delta {\cal N}_k^{\rm SM+\sum Bkg_i} = \sqrt{{\cal N}_k^{\rm SM+\sum_i Bkg_i} \left( 1 + \delta_{\rm sys}^2 \,\,{\cal N}_k^{\rm SM+\sum_i Bkg_i} \right)}.
\end{eqnarray}
The $\chi^2$ analysis due to un-correlated systematic uncertainties
is studied for three representative values of $\delta_{\rm sys}$ at 1\%, 5\%
and 10 \%, respectively. And the sensitivity of $\left\vert V_{tb}\right\vert
\Delta f_1^L$ at 95\% C.L. is found to be of the order of $\sim 10^{-3} - 10^{-2}$
with the corresponding variation of 1\% - 10\% in the systematic error (which
includes the luminosity error). The order of the sensitivity for other anomalous
couplings varies between $\sim 10^{-2}- 10^{-1}$ at 95 \% C.L. 

\subsubsection{Errors and correlations:}
Further, defining the combined $\chi^2_{comb.}\left(f_i,f_j\right)$ and taking
into account of luminosity error $L\equiv \beta \bar L, \beta = 1 \pm \Delta\beta$:
\begin{eqnarray}
  \chi^2_{\rm comb.}(f_i,f_j) \rightarrow \chi^2_{\rm comb.}(f_i,f_j,\,\beta ) \equiv  \sum_{k=1}^m\sum_{i=0}^n\sum_{j=0}^n (f_i -\bar{f_i}) 
   \left[ V^{-1} \right]_{ij}^k (f_j - \bar{f_j}) + \left(\frac{\beta_k -1}{\Delta\beta_k}\right)^2,
  \label{chisq_fifj_lum}
\end{eqnarray}
the sensitivity of $\left\vert V_{tb}\right\vert \Delta f_1^L\sim 10^{-2}$ and
for other couplings it is $\sim 10^{-4}$ for $\Delta \beta \ge 5\%$.

Our analysis shows that the anomalous $Wtb$ vertex at the LHeC can be probed to
a very high accuracy in comparison to all existing limits.

\section{The Higgs boson at the LHeC}
\label{Higgslhec} 
As mentioned in the introduction, the Higgs boson searches are of utmost importance
for all past and present colliders. However, some properties need
to be measured accurately and in this respect future colliders also play a very important
role. In Ref.~\cite{Han:2009pe} the authors studied the $hb\bar b$ coupling at the LHeC,
and they demonstrated that the requirement of forward jet tagging in charged
current events strongly enhances the signal-to-background ratio. The charged
current process at the LHeC is $W$-vector boson fusion and hence
one could measure $hWW$ coupling strength as well. CP properties of the Higgs boson
can be determined by considering an effective five-dimensional vertex, given as
\cite{Biswal:2012mp},
\begin{equation}
\Gamma_{\mu\nu}\left(p,q\right) = \frac{g}{M_W}\left[\lambda\left(p\cdot q g_{\mu\nu} - p_\nu q_\mu \right) + i \lambda^\prime \epsilon_{\mu\nu\rho\sigma} p^\rho q^\sigma \right],
\end{equation}
where $\lambda$ and $\lambda^\prime$ are the effective coupling strengths for
the anomalous CP-conserving and the CP-violating operators, respectively. They
have shown that the azimuthal angle between missing energy and non-$b$ jet
$\Delta\phi_{MET-J}$ is a powerful and unambiguous probe of anomalous $hWW$
couplings, both for CP-conserving and violating type. 

\section{Double Higgs boson production at the FCC-he}
\label{Higgsfcc} 
Further plans for a high-energy LHC provides 50 TeV protons and hence the LHeC center
of mass energy could be increased up  to $\sim 3.5$ TeV with 60 GeV electrons, 
named as the FCC-he. This energy provides opportunities to probe the Higgs self 
coupling strength $g^{(1)}_{hhh}$ through double Higgs boson production.
In this work we consider the speculated detector parameters and cut-based analysis
to get charged-current signals with respect to all possible charged/neutral-current
and photo-production backgrounds~\cite{muk:2015dh}. A statistical analysis is also
performed to find the sensitivity of $g^{(1)}_{hhh}$ with other effective couplings
described with the following effective Lagrangian:
\begin{align}
 {\cal L}^{(3)}_{hhh} =&\,   \frac{m^2_h}{2v} (1 - g^{(1)}_{hhh}) h^3
 + \frac{1}{2} g^{(2)}_{hhh} h \partial_\mu h \partial^\mu h,  \label{laghh1}\\
 {\cal L}^{(3)}_{hWW} =& - \frac{g}{2m_W} g^{(1)}_{hWW} W^{\mu\nu} W^\dag_{\mu\nu} h
 - \frac{g}{m_W} \left[ g^{(2)}_{hWW} W^\nu \partial^\mu W^\dag_{\mu\nu} h + {\rm h.c.} \right] \notag\\
 & - \frac{g}{2m_W} \tilde g_{hWW} W^{\mu\nu} \tilde W^\dag_{\mu\nu} h,  \label{laghh2}\\
 {\cal L}^{(4)}_{hhWW} =& - \frac{g^2}{4m^2_W} g^{(1)}_{hhWW} W^{\mu\nu} W^\dag_{\mu\nu} h^2
 - \frac{g^2}{2m^2_W} \left[ g^{(2)}_{hhWW} W^\nu \partial^\mu W^\dag_{\mu\nu} h^2 + {\rm h.c.} \right]  \notag \\
 & - \frac{g^2}{4m^2_W} \tilde g_{hhWW} W^{\mu\nu} \tilde W^\dag_{\mu\nu} h^2.  \label{laghh3}
\end{align}
Here $g^{(1)}_{hhh}$ is defined such that it appears as a multiplicative constant
to $\lambda_{SM}$ i.e; $\lambda \to g^{(1)}_{hhh}\lambda_{SM}$ in the potential
for electroweak symmetry breaking in the SM:
\begin{align}
 V(\Phi)
 = \mu^2 \Phi^\dag \Phi + \lambda (\Phi^\dag \Phi)^2
 \to \frac{1}{2}m^2_h h^2 + \lambda v h^3 + \frac{\lambda}{4} h^4,
 \label{Vphi}
\end{align}
with $\lambda = \lambda_{\rm SM} = m^2_h/(2v^2) \approx 0.13$. The effective
couplings  $g^{(1)}_{hhh}$, $g^{(2)}_{hhh}$, $g^{(1)}_{hWW}$, $g^{(2)}_{hWW}$,
$g^{(1)}_{hhWW}$, and $g^{(2)}_{hhWW}$ are CP-conserving,
whereas $\tilde{g}_{hWW}$, $\tilde{g}_{hhWW}$ are CP-violating effective couplings. 
$m_h$ and $m_W$ are respectively masses of the Higgs and $W$-bosons, 
$W^{\mu\nu} = \partial^\mu W^\nu - \partial^\nu W^\mu$.

\subsection{Cross section, Detector setup and cut-based analysis}
Fiducial cross sections for signal and backgrounds, before cut-based analysis,
are shown in
Table~\ref{xsec}. For signal we consider the charged current process $p e^- \to
\nu_e h h j, h \to b \bar b$. Photo-production backgrounds are very important for
signals, other charged/neutral-current backgrounds, and those backgrounds
estimated through ``Equivalent photon approximation structure functions''; which
is calculated with the ``Improved Weizsaecker-Williams formula'' \cite{Budnev:1974de}.
\begin{table}[tb]\footnotesize
\centering
\begin{tabular}{|lccc|}\hline\hline
Process                    & {\scshape {cc}} (fb)  & {\scshape {nc}} (fb)  & {\scshape{photo}} (fb) \\ \hline\hline
Signal:                    & $2.40 \times 10^{-1}$ & $3.95 \times 10^{-2}$ &  $3.30 \times 10^{-6}$ \\ \hline\hline
$b\bar b b\bar b j$:       & $8.20 \times 10^{-1}$ & $3.60 \times 10^{+3}$ &  $2.85 \times 10^{+3}$ \\ \hline
$b\bar bjjj$:              & $6.50 \times 10^{+3}$ & $2.50 \times 10^{+4}$ &  $1.94 \times 10^{+6}$ \\ \hline
$zzj$($z\to b\bar b$):     & $7.40 \times 10^{-1}$ & $1.65 \times 10^{-2}$ &  $1.73 \times 10^{-2}$ \\ \hline
$t\bar tj$(hadronic):      & $3.30 \times 10^{-1}$ & $1.40 \times 10^{+2}$ &  $3.27 \times 10^{+2}$ \\ \hline
$t\bar tj$(semi-leptonic): & $1.22 \times 10^{-1}$ & $4.90 \times 10^{+1}$ &  $1.05 \times 10^{+2}$ \\ \hline
\end{tabular}
\caption{\small {Cross sections (in fb): $E_e = 60$ GeV, $E_p = 50$ TeV, $j = g u \bar u d \bar d s \bar s  c \bar c$. Initial cuts: $|\eta| \le 10$ for jets, leptons and $b$, $P_T \ge 10$ GeV, $\Delta R_{\rm min} = 0.4$ for all particles. 
}}
\label{xsec}
\end{table}

In the detector setup, the maximum rapidity $\left|\eta \right|$ range is up to 7.
For the $b$-tagging, the jets with $\left| \eta\right|<5$ and transverse momentum
$p_T > 15$ GeV is taken. The fake rate for a $c$-initiated jet and a light jet 
to the $b$-jet is 10\% and 1\% respectively. The weight corresponding to the 
$b$-tagging efficiency or fake rate is assigned to each event. Furthermore, the 
following cut flows are taken for analysis:
\begin{itemize}
\item Select 4 $b$ + 1-jet: $p_T^{jet} > 20$ GeV, $|\eta|<7$ for $non$-$b$-jets,
$|\eta|<5$ for b-jets. The four $b$ jets must be well separated within $\Delta R>0.7$
\footnote{The $\Delta R$ is defined as the distance between two objects $i$ and $j$ in the 
rapidity-azimuthal plane: $\Delta R = \sqrt{\left( \phi_i-\phi_j \right)^2 + \left(\eta_i-\eta_j \right)^2}$,
where $\phi_i$ and $\eta_i$ are the azimuthal angle and the rapidity of the object $i$.}
in case of overlapped truth matching in the $b$-tagging.

\item Rejecting leptons with $p_T^{e^-} > 10$ GeV (to suppress the neutral-current
process).

\item $\eta_{forward-jet} > 4.0$, the forward jet as defined as the $non$-$b$-jet
which has the largest $p_T$ after selecting at least 4 $b$-jets.

\item $E_{T}^{miss} > 40$ GeV and $\Delta\Phi_{E_{T}^{miss}, leadingjet} > 0.4$,
$\Delta\Phi_{E_{T}^{miss}, subleadingjet} > 0.4$ \footnote{$\Delta \Phi_{ij}$ is 
the azimuthal angle difference of two objects $i$ and $j$. The
(sub)leading jet is the $non$-$b$ jet defined with (second-)highest $p_T$.}.

\item Pair the four $b$-jets into two pairs and calculate the invariant masses
of each pair. The composition of the pair which have the smallest variance of
mass to $(m_H - 40)$ GeV is chosen. The first pair is defined as $90<M_{1}<125$ GeV,
which must have the leading $b$-jet. The other pair is defined as $75<M_{2}<125$ GeV.

\item Choosing the invariant mass of all four $b$-jets greater than 280 GeV.  
\end{itemize}

And the significance is calculated with a Poisson distribution \footnote{$s=\sqrt{2
\left(\left(S+B\right) log \left(1+S/B \right)-S \right)}$}, considering the
expected signal (S) and background (B) yields at 10 ${\rm ab}^{-1}$ luminosity. 
After performing these cut based analyses the signal events are $\sim 63$ with 
respect to total background events $\sim 35$ and $s = 8.7$. 
A 5\% systematic error is introduced into signal and backgrounds.

\begin{figure*}[tb]
  \centering
  \subfloat[]{\includegraphics[width=0.5\textwidth,clip]{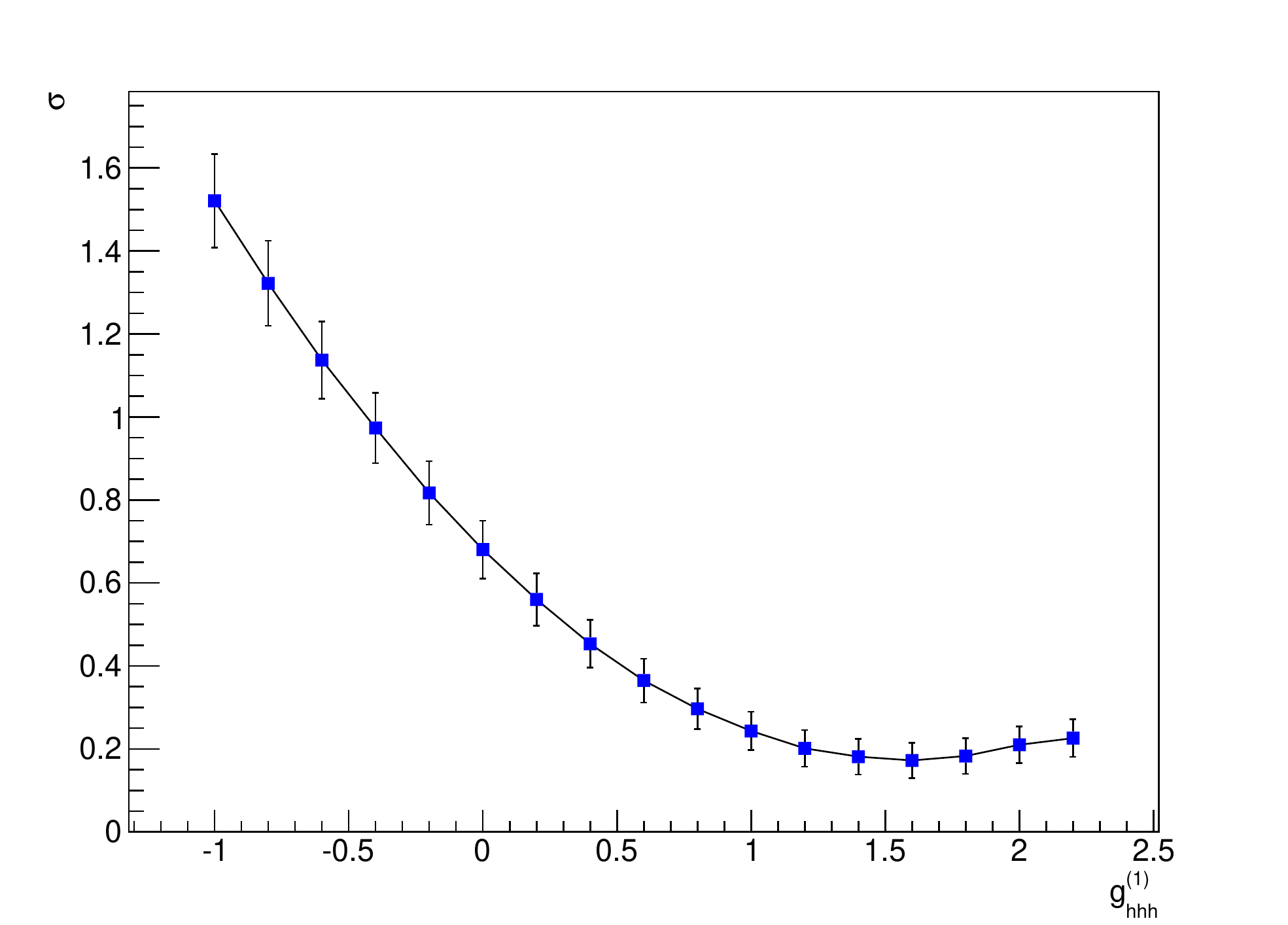}} 
  \subfloat[]{\includegraphics[width=0.5\textwidth,clip]{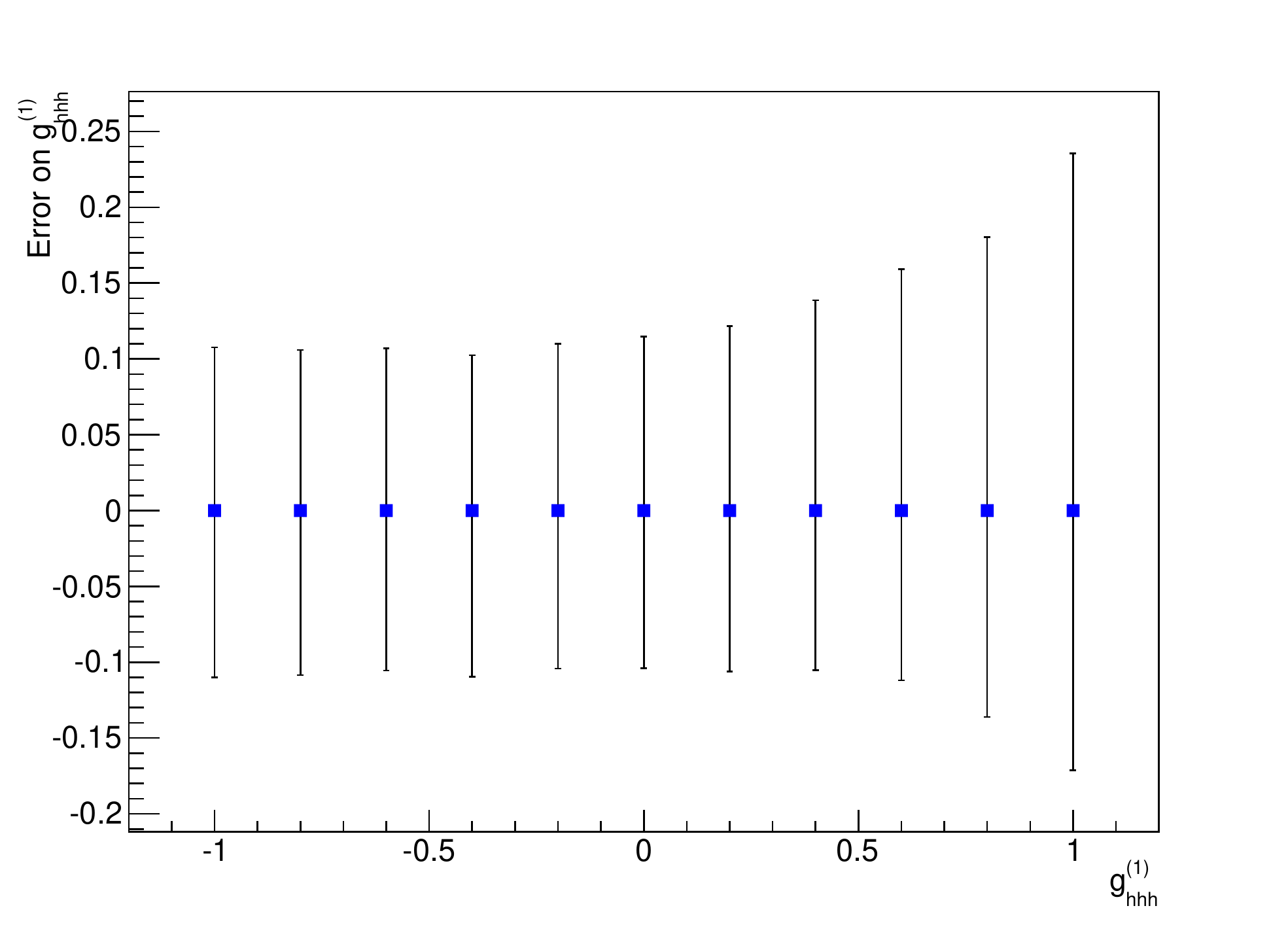}} 
  \caption{\small (a) Variation of cross section of the signal process
  $\sigma(p e^- \to \nu_e h h j, h \to b \bar b)$ with respect to $g^{(1)}_{hhh}$
  with error bar at each value of $g^{(1)}_{hhh}$, (b) local error through linear
  interpolation at each value of $g^{(1)}_{hhh}$. }
\label{fig1}
\end{figure*}

\subsection{Statistical analysis}  
\begin{figure*}[tb]
  \centering
   \subfloat[]{	\includegraphics[width=\textwidth,height=0.42\textwidth]{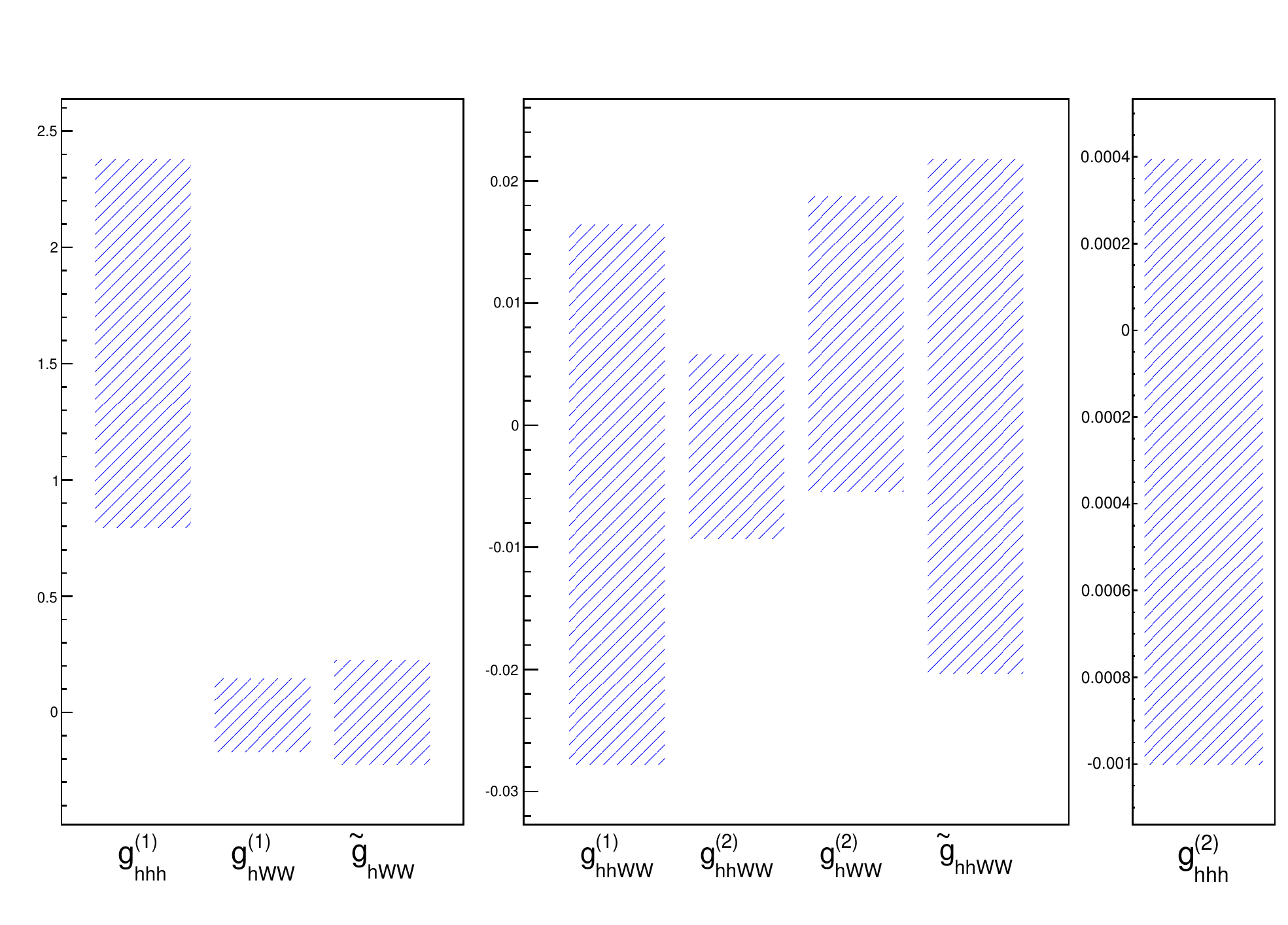}}
    \caption{\small The limits on the coupling strength, derived at 0.4 ${\rm ab}^{-1}$.
    The $g^{(2)}_{hhh}$ has only the upper limit because the cross section dependence is
    monotonic in this region.}
\label{fig2}
\end{figure*}

Following the method given in Ref.~\cite{Cowan:2010js}, exclusion limits for
$g^{(1)}_{hhh}$ are calculated. Fig.~\ref{fig1} shows significant behaviour
of cross section variation with respect to $g^{(1)}_{hhh}$, which is expected,
due to interference between resonant and non-resonant Higgs mediation in the
charge-current signal process. The 95\% upper limits of all effective couplings
appear in the Lagrangian Eqs. (\ref{laghh1}), (\ref{laghh2}) and (\ref{laghh3}) 
due to cross section influence, are also calculated and shown in Fig.~\ref{fig2}. 
The sensitivity of CP-even and odd $HWW$ effective couplings, $g^{(1)}_{hWW}$ and
$\tilde g_{hWW}$, are of the same order as $g^{(1)}_{hhh}$ $\sim 10^0$. However, 
for $g^{(2)}_{hhh} \sim 10^{-3}-10^{-4}$ and $g^{(2)}_{hWW}$, $g^{(1,2)}_{hhWW}$,
$\tilde g_{hhWW}$, these are of the order $\sim 10^{-2}$.

\section{Conclusion}
\label{conc}
We briefly reviewed the physics potential of future $e^- p$ colliders, speculated
to build on top of the LHC, through the top quark and Higgs boson physics.
And we infer that the LHeC and the FCC-he is a viable option to study top 
and Higgs physics, and for the measurement of related couplings to high accuracy.  

\section*{Acknowledgements}
MK would like to acknowledge all his collaborators, namely Bruce Mellado, Sukanta
Dutta, Ashok Goyal, Alan Cornell, Xifeng Ruan and Rashidul Islam, as well as the organisers
of the HEPP workshop 2015.


\section*{References}
\begin{thebibliography}{9}

\bibitem{Dutta:2013mva} 
  S.~Dutta, A.~Goyal, M.~Kumar and B.~Mellado,
  arXiv:1307.1688 [hep-ph].

\bibitem{Han:2009pe} 
  T.~Han and B.~Mellado,
  Phys.\ Rev.\ D {\bf 82}, 016009 (2010)
  [arXiv:0909.2460 [hep-ph]].

\bibitem{Biswal:2012mp} 
  S.~S.~Biswal, R.~M.~Godbole, B.~Mellado and S.~Raychaudhuri,
  Phys.\ Rev.\ Lett.\  {\bf 109}, 261801 (2012)
  [arXiv:1203.6285 [hep-ph]].

\bibitem{muk:2015dh}
 In preparation: M.~Kumar, X.~Ruan, A.~Cornell, R.~Islam, M.~Klein, U.~Klein, B.~Mellado

\bibitem{Budnev:1974de}
  V.~M.~Budnev, I.~F.~Ginzburg, G.~V.~Meledin and V.~G.~Serbo,
  Phys.\ Rept.\  {\bf 15} (1975) 181.
 
\bibitem{Cowan:2010js}
  G.~Cowan, K.~Cranmer, E.~Gross and O.~Vitells,
  Eur.\ Phys.\ J.\ C {\bf 71} (2011) 1554
   [Erratum-ibid.\ C {\bf 73} (2013) 2501]
  [arXiv:1007.1727 [physics.data-an]].
 

\end{thebibliography}
\end{document}